\documentclass[aps,twocolumn,amsmath,letterpaper,floatfix]{revtex4}
\usepackage{amssymb}
\usepackage{graphicx}
\usepackage{array}
\usepackage{hhline}
\usepackage{longtable}
\usepackage{float}

\begin{document}

\title{Chaotic Spin Correlations in Frustrated Ising Hierarchical Lattices}

\author{Ne\c{s}e Aral$^1$ and A. Nihat Berker$^{1-3}$}
\affiliation{$^1$College of Arts and Sciences, Ko\c{c} University,
Sar\i yer 34450, Istanbul, Turkey,} \affiliation{$^2$Department of
Physics, Massachusetts Institute of Technology, Cambridge,
Massachusetts 02139, U.S.A.,} \affiliation{$^3$Feza G\"ursey
Research Institute, T\"UB\.ITAK - Bosphorus University,
\c{C}engelk\"oy 34684, Istanbul, Turkey}

\begin{abstract}
Spin-spin correlations are calculated in frustrated hierarchical
Ising models that exhibit chaotic renormalization-group behavior.
The spin-spin correlations, as a function of distance, behave
chaotically. The far correlations, but not the near correlations,
are sensitive to small changes in temperature or frustration, with
temperature changes having a larger effect.  On the other hand, the
calculated free energy, internal energy, and entropy are smooth
functions of temperature.  The recursion-matrix calculation of
thermodynamic densities in a chaotic band is demonstrated.  The
leading Lyapunov exponents is calculated as a function of
frustration.


\end{abstract}

\maketitle
\def\s{\rule{0in}{0.28in}}

It was shown some time ago that frustrated Ising spin magnetic
systems exhibit chaotic behavior of the interaction constants under
renormalization-group transformations, which readily leads to the
description of a spin-glass phase.\cite{McKayBerkerKirk} This
chaotic rescaling behavior was originally demonstrated in
frustrated, but non-random systems.  It was subsequently shown that
the same chaotic rescaling behavior occurs in quenched random spin
glasses.\cite{McKay} Chaotic rescaling behavior has now been
established as the signature of a spin-glass phase.\cite{Krzakala,
Marinari, LeDoussal, Rizzo, Sasaki,
Bouchaud,Yoshino,Aspelmeijer1,Aspelmeijer2,Mora} Although the
chaotic behavior of the interaction constants was demonstrated in
frustrated systems, the behavior of spin-spin correlation functions
and the instabilities to initial conditions had not been calculated
to-date.  This study presents such results, yielding both smooth and
unsmooth behaviors, as seen below.  In this process, the
recursion-matrix calculation of thermodynamic densities in a chaotic
band is demonstrated.

\begin{figure}[h]
\includegraphics*[scale=1]{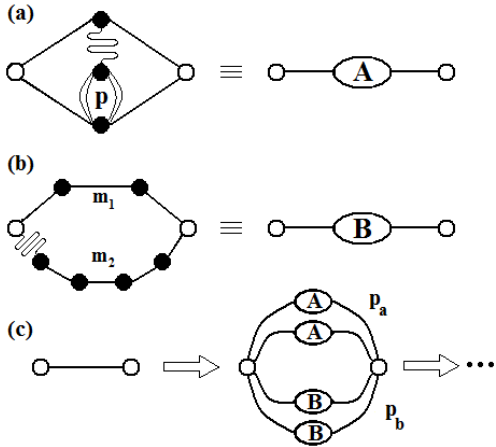}
\caption{The family of hierarchical models from
Ref.\cite{McKayBerkerKirk}.  In unit (a), there are $p$ cross bonds.
In unit (b), two paths, consisting of $m_1$ and $m_2>m_1$ bonds in
series, are in parallel.  In (c), the final graph of the model is
assembled with $p_a$ and $p_b$ of each unit in parallel. Each wiggly
bond, representing an infinite antiferromagnetic coupling, has the
effect of reversing the sign of an adjoining bond.} \label{model}
\end{figure}

\begin{figure}[h]
\includegraphics*[scale=1]{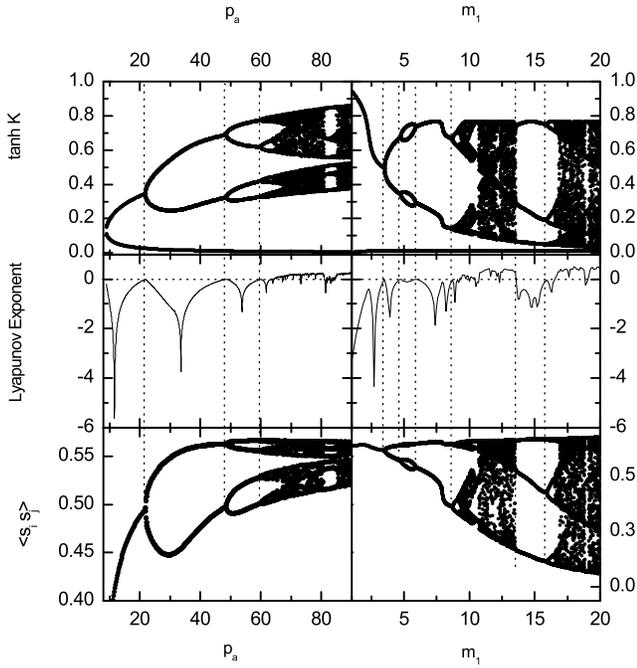}
\caption{Renormalization-group flow topologies, Lyapunov exponents,
and spin-spin correlations, as Hamiltonian parameters are scanned.
In both examples, $p=5, p_b=2, m_2=m_1+5$.  In the upper panels, the
lower curve, visible on the left, is a line of unstable fixed
points, giving the second-order phase transition between the
paramagnetic (below) and ordered (above) phases.  In the ordered
phase, only fixed points, limit cycles, and chaotic bands that are
stable (attractive) are shown. In the lower panels, the spin-spin
correlation function $<s_is_j>$ for spins separated by a distance
$2^n$ are given for consecutive $n$ at each value of the Hamiltonian
parameters. It is shown in the text that the Lyapunov exponents,
middle panels, apply to both upper and lower panels.  Left panels:
Scanning $p_a$, which increases the \textit{disordering by ordering}
effect, at $m_1=5$. Right panels: Scanning $m_1$, which increases
the ground-state entropy of per first renormalized bond, at
$p_a=50$.} \label{flow}
\end{figure}

Hierarchical models are exactly soluble models that exhibit
non-trivial cooperative and phase transition behaviors
\cite{BerkerOstlund,Kaufman,Kaufman2} and have therefore become the
testing grounds for a large variety of phenomena, as also seen in
recent works.\cite{Erbas,Hinczewski2,Hinczewski3,ZZC,
Khajeh,Rozenfeld,Kaufman3,Piolho,Branco,Jorg,Boettcher,Monthus}. The
hierarchical models in which the chaotic rescaling behavior of the
interaction constants under frustration was seen
\cite{McKayBerkerKirk} are defined in Fig.\ref{model}.  The two
units [Fig.\ref{model}(a) and (b)] assembled in the construction of
these lattices \textit{a priori} represent the generically distinct
local effects of frustration occurring in spin-glass systems on
conventional lattices:  In Fig.\ref{model}(a), correlation at the
small scale (vertical bonds) inhibit at low temperatures the
propagation of correlation at the larger scale (horizontally across
the unit), namely causing \textit{a disordering by ordering}. In
Fig.\ref{model}(b), competition between paths of different lengths
weakens but does not eliminate the propagation of correlation across
the unit.  These two generic effects are incorporated into the
hierarchical lattices of Fig.\ref{model}. No other such generic
effects occur in spin glasses.

\noindent\textit{Renormalization-Group Transformation.} Hierarchical
lattices are constructed by the repeated self-imbedding of
graphs.\cite{BerkerOstlund,Kaufman,Kaufman2} Their solution, by
renormalization-group theory, consists of the reverse procedure. The
number of bonds of the imbedding graph gives the volume rescaling
factor, $b^d = (4+p)p_a + (m_1 + m_2)p_b$ in the current case, and
the shortest path length across the imbedding graph gives the length
rescaling factor, $b = 2$ here, leading here to a dimensionality $d$
that is greater than 2.  Each straight line segment in
Fig.\ref{model} corresponds to an interaction $ - \beta H_{ij} =
K\sigma_{i}\sigma_{j} + G$ with $K\geq0$ between Ising spins
$\sigma_{i} = \pm 1$ at vertices $i$. Frustration is introduced by
the wiggly bonds.  The additive constant $G$ is generated by the
renormalization-group transformation and enters the calculation of
the thermodynamic functions and correlations of the original,
unrenormalized system.\cite{McKayBerker}.  The renormalization-group
transformation consists in summing, in the partition function, over
the internal spins of the innermost imbedding graphs, which are
thereby replaced by a renormalized bond with
\begin{multline}
K' = p_a \tanh^{-1}\tilde{t}_a + p_b(\tanh^{-1}t^{m_1} -
\tanh^{-1}t^{m_2}),\\
\text{where }\quad \tilde{t}_a = 2 t^2 (1 - \tilde{t})/(1 + t^4 - 2 t^2\tilde{t}),\\
G' = b^dG + [2 p_a + (m_1 + m_2 - 2)p_b]\ln 2 - p_a p K\\
+ \frac{p_a}{2} \ln \frac{(1 + t^2)^2 - 4 t^2 \tilde{t}}{(1 - t^2)^2
(1 -\tilde{t})^2} + \frac{p_b}{2} \ln \frac{(1 -t^{2 m_1})(1 - t^{2
m_2})}{(1 - t^2)^{m_1 + m_2}},
\end{multline}
with $t = \tanh K$ and $\tilde{t} = \tanh(pK)$.

\begin{figure}[h]
\includegraphics*[scale=0.9]{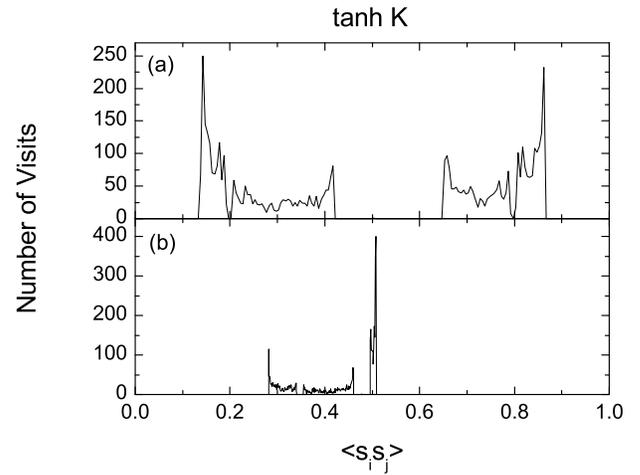}
\caption{(a) Number of visits per interaction interval $\Delta t$ =
0.005, for 5000 chaotic iterations in the trajectory starting at
$t^{(0)} = 0.5$, for $p=4, p_a=40, p_b=1, m_1=4.7, m_2=m_1+1$. (b)
Number of visits per correlation interval $\Delta <s_is_j>$ = 0.005
for the trajectory in (a) } \label{nov}
\end{figure}

\begin{figure}[!h]
\includegraphics*[scale=1]{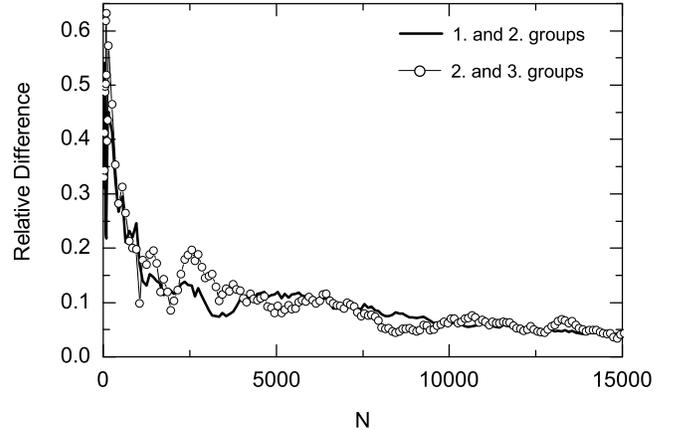}
\caption{Overlaps between consecutive groups of N iterations for the
trajectory in Fig.3(a).  The relative difference
$\frac{1}{200}\Sigma_{i=1}^{200} |\Delta n_i| / \bar{n}_i$,  between
two consecutive groups, in the number of visits $n_i$ to each of the
200 interaction intervals $i$ is shown as a function of group size
$N$.} \label{overlap}
\end{figure}

\begin{figure}[!h]
\includegraphics*[scale=1]{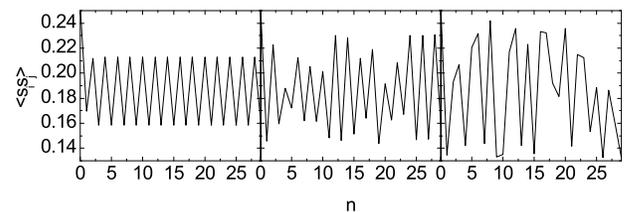}
\caption{The spin-spin correlation function $<s_is_j>$ for spins
separated by a distance $2^n$, for $K = 2.5, p=4, p_a=40, p_b=1, m_1
= 3.7,4.7,5.7, m_2=m_1+1$.} \label{chaos}
\end{figure}

\begin{figure}[!h]
\includegraphics*[scale=1]{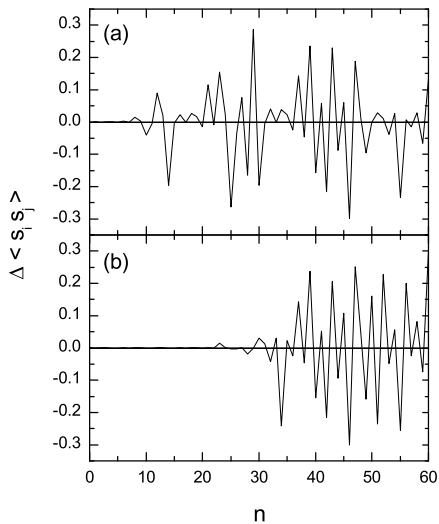} \caption{Deviations,
for small temperature or frustration change, in the spin-spin
correlation function $<s_is_j>$ for spins separated by a distance
$2^n$, for $K =0.8, p=4, p_a = 40, p_b=1, m_1 = 8, m_2=m_1+1$. In
(a) and (b), between the two trajectories, $\Delta K = 0.001$ and
$\Delta p_a = 0.001 $ respectively.} \label{chaotic_b}
\end{figure}

\begin{figure}[h]
\includegraphics*[scale=0.8]{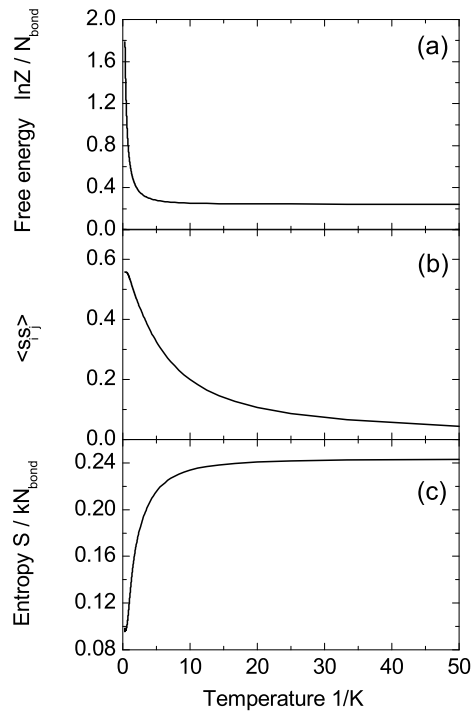} \caption{For model
parameters $p=4, p_a = 10, p_b=1, m_1 = 7, m_2=m_1+1$, (a) free
energy per bond, $F = \ln Z / N_{bond}$, (b) internal energy, $U =
<s_is_j>$, (c) entropy per bond, $S = \ln Z / N_{bond} - K
<s_is_j>$, versus temperature $K^{-1}$.} \label{U}
\end{figure}

\noindent\textit{Results: Self-Similar Chaotic Bands.} When this
family of models is scanned as a function of $p$ or $m_1$,
respectively increasing the \textit{disordering by ordering} effect
or the ground-state entropy per first renormalized bond, the chaotic
behavior of the renormalization-group is entered, in the
low-temperature phase, via the series of period-doubling
bifurcations, with the Feigenbaum exponent of
4.669~\cite{McKayBerkerKirk}, as shown in Fig.\ref{flow}. An example
of the chaotic bands of the interaction constant is shown in
Fig.\ref{nov}(a).  Discovery of these chaotic bands immediately led
to a spin-glass interpretation:  Under repeated scale changes, the
entire band is visited by the effective coupling of the length
scales that are reached after each renormalization-group
transformation.  This chaotic sequence of hopping visits stretches
from the strong coupling to the weak coupling edges of the band.
This signifies that, as the system is viewed at successive length
scales, strong and weak correlations are encountered in a frozen but
chaotic sequence, meaning a spin-glass phase.  This interpretation
had not been followed by an actual calculation of these spin-spin
correlations, which is done in the current study.

Also shown in Fig.\ref{flow} are the leading Lyapunov exponents
($\lambda$), used to describe the behavior of a dynamical system
that starts at $x_0$ and evolves for $n$ iterations, $x_{i+1} =
f(x_i)$,
\begin{equation} \label{eq:2a}
\lambda =  \begin{array}{c} \lim \\ n \to \infty \\ \end{array} \,
\frac{1}{n} \sum_{k=0}^{n-1} \ln|\frac{d x_{k+1}}{d x_k}|.
\end{equation}
The iteration function $f$ may depend on different parameters, as
our iteration function $K'(K)$ depends on $m_1$ and $p_a$.  Such an
iterated map function has a chaotic trajectory for a particular
parameter value if the Lyapunov exponent is positive.  Conversely, a
negative $\lambda$ indicates eventual attraction to a fixed point or
a limit cycle.  A bifurcation point, where a period doubling occurs,
is identified with $\lambda$ being zero.\cite{Lyapunov} Furthermore,
\begin{multline}\label{eq:3a}
\frac{d <s_is_j>_{k+1}}{d <s_is_j>_{k}} =\\
\frac{d <s_is_j>_{k+1}}{d \tanh(K_{k+1})}\cdot \frac{d
\tanh(K_{k+1})}{d \tanh(K_k)}\cdot \frac{d \tanh(K_{k})}{d
<s_is_j>_{k}}\,,
\end{multline}
so that the first and last factors from Eq.~\eqref{eq:3a} cancel out
in the successive terms in Eq.~\eqref{eq:2a}.  Thus, the interaction
constants and the spin-spin correlations have the same Lyapunov
exponents.

An important characteristic of the chaotic bands is that they are
self-similar:  After the transient behavior of a number of
renormalization-group transformations, the profile of the chaotic
band formed by each successive group of N renormalization-group
calculations becomes identical in the limit of large N.  The
overlaps between such successively formed bands is shown as a
function of N in Fig.4. Physically, this signifies that a
geometrically coarse-grained spin-glass phase is self-similar.  This
property of the chaotic bands is important in the calculation of the
correlation functions.

\noindent\textit{Results: Calculation of the Correlation Functions.}
The recursion relations for the densities is \cite{McKayBerker}
\begin{equation}
[1, <s_i s_j>] = b^{-d}[1, <s_i s_j>'] \left( \begin{array}{ccc}
                b^d & \frac{\partial G^\prime}{\partial K} \\
                0 & \frac{\partial K^\prime}{\partial K} \\
            \end{array}\right)\ .
\end{equation}
In an ordinary renormalization-group analysis, this density
recursion relation is iterated,
\begin{equation} \label{eq:5}
[1, <s_i s_j>] = b^{-dn}[1, <s_i s_j>^{(n)}] \cdot T ^{(n)}\cdot T
^{(n-1)}\cdot...\cdot T ^{(1)},
\end{equation}
until the $(n)$th renormalized system is as close as one desires to
a sink fixed point, and the renormalized densities $[1, <s_i
s_j>^{(n)}]$ are inserted as the left eigenvector of $T ^{(n)}$ with
eigenvalue $b^d$.\cite{McKayBerker}  In the current calculation,
this cannot be done, since the renormalization-group trajectory does
not approach a sink fixed point, but chaotically wanders inside a
band.  On the other hand, in this chaotic-band sink, we obtain the
limiting behavior, due to the self-similarity property of the
chaotic band,
\begin{equation}
b^{-dn} T ^{(n)}\cdot T ^{(n-1)}\cdot...\cdot T ^{(1)} \simeq \left(
\begin{array}{ccc}
                1 & X \\
                0 & 0 \\
            \end{array}\right)\ ,
\end{equation}
so that $X=<s_i s_j>$ and this result is independent of the
chaotic-band terminus $<s_i s_j>^{(n)}$.  The disappearance of the
lower diagonal reflects $\partial K^\prime/\partial K < b^d$, itself
due to sequential non-infinite bonds and frustration in the
chaotic-band sink.  Alternately, $<s_i s_j>$ can be calculated from
numerical differentiation of the free energy obtained from the
renormalization of the additive constant $G$.

The calculated spin-spin correlations as a function of spin
separation are shown in Fig.\ref{chaos}.  It is seen that the
spin-spin correlations behave chaotically, for all distances,
between strong and weak correlations, numerically justifying the
spin-glass phase interpretation.  Thus, spin-spin correlations also
span chaotic bands, as illustrated in Fig.2, lower panels, and Fig.
\ref{nov}(b).

\noindent\textit{Results: Unsmooth and smooth behaviors.}
Fig.\ref{chaotic_b} shows the behavior, at all distances, of the
spin-spin correlations under small changes in temperature or
frustration.  It is seen that the near correlations are unaffected,
whereas the far correlations are strongly affected, namely randomly
changed, with temperature changes having a larger such effect.

Finally, the free energy, calculated from summing the additive
constants generated by the successive renormalization-group
transformations, the internal energy, calculated from the
nearest-neighbor spin-spin correlation, and the entropy are shown in
Figs.\ref{U}(a-c) as a function of temperature.  They exhibit smooth
behaviors.  Zero-temperature entropy \cite{BerkerKadanoff}, due to
frustration, is seen.

In closing, we note that other forms of chaotic behavior, namely as
a function of system size \cite{Newman} or as the chaos of
near-neighbor correlations in the zero-temperature limit for
appropriately chosen interactions \cite{vanEnter}, intriguingly
occur in spin-glass systems.  In contrast to our current results,
the renormalization of correlations in a strange non-chaotic
attractor are given in \cite{Politi}.

\begin{acknowledgments}
This research was supported by the Scientific and Technological
Research Council of Turkey (T\"UB\.ITAK) and by the Academy of
Sciences of Turkey.
\end{acknowledgments}

\end{document}